\documentclass[referee]{aa}
\usepackage{amsmath}
\usepackage[dvips]{graphics}
\usepackage {epsfig}
\title{Low surface brightness galaxies mass profiles as a consequence of galactic evolution.}
\author{C. Alard}

\institute{Institut d'Astrophysique de Paris, 98bis boulevard Arago, 75014
           Paris \\
           \email{alard@iap.fr}}
\date{}
\begin{document}
\abstract {} 
{This paper presents a principal components analysis of rotation curves from a sample of low surface brightness
galaxies. The physical meaning of the principal components is investigated, and related
to the intrinsic properties of the galaxies.}
{The rotation curves are re-scaled using the optical disk scale, the resulting principal
component decomposition demonstrates that the whole sample is properly approximated using two components.}
{The ratio of the second to the first component is related to the halo steepness in the central region,
is correlated to the gas fraction in the galaxy, and is un-correlated to other parameters. As a consequence
 the gas fraction appear as a fundamental variable with respect to the galaxies rotation curves, and its 
correlation with the halo steepness is especially important.}
{Since the gas fraction is related to the degree of galaxy evolution, it is very likely that
the steepness of the halo at the center is a consequence of galaxy evolution. More evolved 
galaxies have shallower central profile and statistically less gas, most likely as a consequence
of more star formation and supernovae. The differences in evolution, gas fractions and halo central
steepness of the galaxies could be due to the influence of different environments.}
\maketitle
\section{Introduction.}
Cold dark matter predicts steep density profile at the center
of dark matter halo's (Navarro, Frenk \& White 1996; Navarro, 
Frenk \& White 1997; Moore, {\it et al.}, 1998). This prediction is not 
consistent with observations of constant core density in low surface 
brightness galaxies (Moore, 1994; Burkert, 1995; Navarro, Frenk \& White
1996; McGaugh \& de Blok, 1998; de Blok \&  Bosma 2002). Various modifications 
of the cold dark matter scenario have been proposed to account for this discrepancy, 
collisional dark matter (Carlson, Machacek, \& Hall, 1992; Spergel \& Steinhardt, 2000;
Moore {\it et al.}, 2000), Fluid dark matter (Peebles 2000), Boses Einstein
condensates (Hu, Barkana, \& Gruzinov). Other approaches do not attempt
to modify the cold dark matter scenario, but emphasize the role of supernovae,
ram pressure stripping or tidal interactions as mechanism affecting the mass
profile of dwarf galaxies (Read \& Gilmore 2005). This paper will not attempt
to favor one or another scenario, but will focus on analyzing rotation
curves data without preconceived hypothesis. The approach will focus
on isolating the fundamental features in the rotation curves and will 
relate them to the intrinsic physical parameters of low surface 
brightness galaxies (LSB). 
\section{Data analysis.}
de Blok \& Bosma (2002), (hereafter dBB) published the rotation curves of 
26 LSB. The dBB sample contains galaxies of different sizes and masses requiring the
re-scaling of the rotations curves. In practice, the rotation curves were re-scaled using the length scale 
inferred from the visible light (Table 1 in dBB). Provided
the various rotation curves are related to the same universal physical process, 
the re-scaled curves should be reducible to a minimal set of fundamental components.
 The reduction to a minimal set is equivalent to a
 principal component analysis (PCA). The PCA requires a re-sampling of the curves to a 
common grid, this re-sampling was performed using B-spline interpolation. The PCA
vectors are re-constructed in the range $R_{pc}$  (in re-scaled units). The range $R_{pc}$
must include a sufficient number of points (a minimum of 5 points was required). An additional
requirement was the availability of an estimation of the HI contribution to the rotation curve. The contribution
of the gas was subtracted to the mass budget.
The range $R_{pc}$ was adjusted to be as large as possible while maximizing the number of curves.
 The optimal value is $R_{pc}=2.85$, which corresponds to a total of 15 curves. Since
the mass scale like the square of the velocity the PCA is performed on the 
square of the velocity data. Finally the amplitudes of the curves were re-normalized so that their self
cross product is unity. The PCA matrix is constructed by filling the matrix columns with the rotation curve
vectors. PCA requires to find the eigenvalues of the covariance matrix, numerically
this is best achieved by using a singular value decomposition of the matrix. Components are sorted
by descending order of power spectrum contribution (see Fig. ~\ref{pca1}). 
\subsection{Effect of noise on the PCA analysis.}
\begin{figure}
\centering{\epsfig{figure=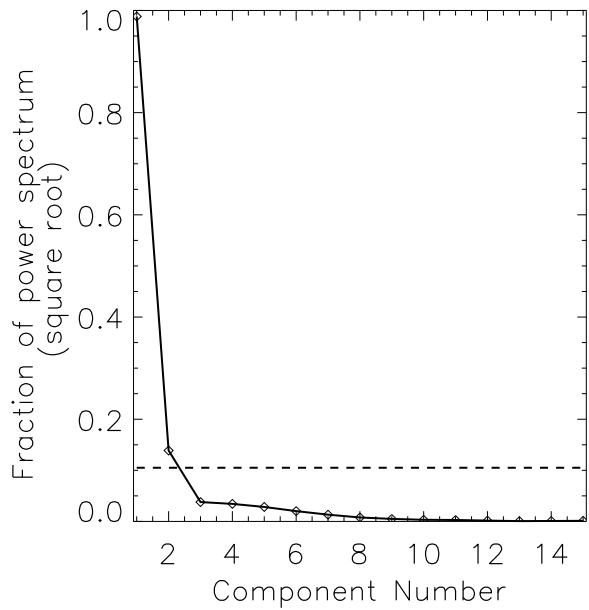,width=9cm}}
\caption{Square root of the power spectrum contribution for each principal
component in order of decreasing contribution. The dotted line represents the 3-$\sigma$
level of the noise expectation.}
\label{pca1}
\end{figure}

\begin{figure}
\centering{\epsfig{figure=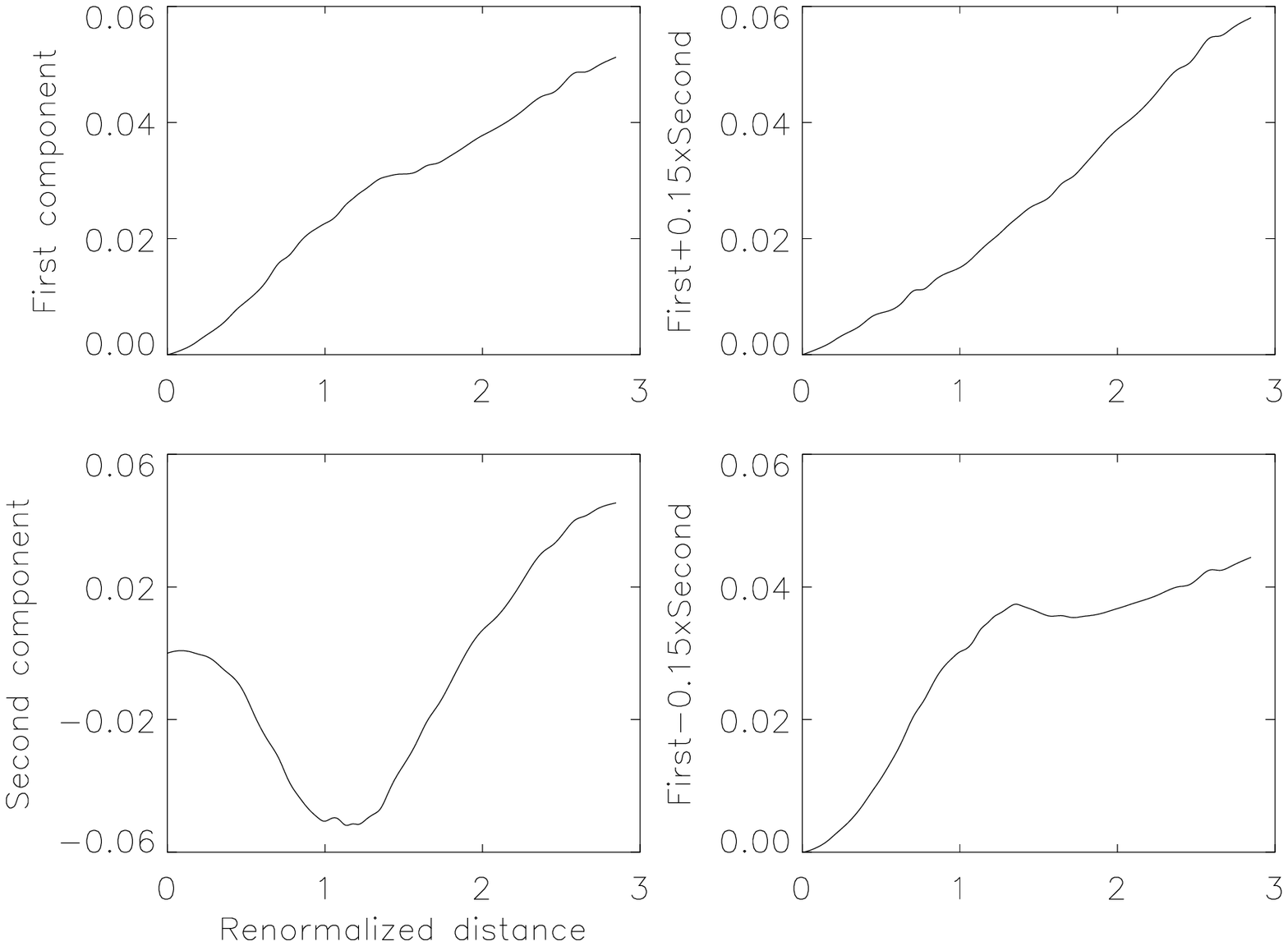,width=13.95cm}}
\caption{The first two components presented together with linear combinations of the components
illustrating the effect of the second component on the rotation curves.}
\label{pca2}
\end{figure}

Numerical experiments
performed using Monte-Carlo simulations with gaussian noise expectations on the velocities (as provided by dBB)
indicates that the amplitude of the noise on the power spectrum of each components (sums of the square
of amplitudes for each vector, normalized by total power), is about $(0.035)^2$. As a consequence
only the first two components are significant (see Fig. ~\ref{pca1}, ~\ref{pca2}). 
Another problem is the stability of these two components with respect to the noise fluctuations 
and the re-scaling errors. Random gaussian noise with variance taken from dBB 
were added to the set of light curves and the singular value decomposition was re-computed each time,
in order to estimate the effect of noise on the principal components. The result of 1000 simulations
shows that the noise accounts for about $0.5\%$ of the total variance on the first component, and $15\%$
of the  total variance on the second component. These numerical experiments shows also that only a linear
interpolation of the rotation curves is stable numerically. Let's 
now investigate the errors introduced by the re-scaling of the rotation curves. 
The curves were re-scaled according to the disk scale length, but there is no guaranty
 that the scale of the other components are exactly
 proportional to the disk scale. To evaluate the effects of uncertainty in the re-scaling,
 uniform random fluctuations of the
scale length were introduced and the PCA was re-conducted each time. The results shows that
for uniform deviations of the scale length with respective amplitudes of, $10\%$, $20$, and $40\%$, 
the corresponding variance of the noise as a fraction of the total variance of the second components is:
$2\%$, $4\%$, and $9\%$. In addition to these results note that the mean correlation 
coefficients between the simulated
second component and the original one is 0.9 when noise fluctuations are considered, and higher
for scale fluctuations. The first components is much less affected and has a mean correlation 
close to unity.
As a conclusion the results on the statistical variance, and the correlation are a good illustration 
of the stability of the PCA analysis of this sample.  
\subsection{Interpretation of the components.}
The first component represents approximately the mean of all curves. 
The meaning of the second component is easily
understood by evaluating its effect on the shape of the curves. Adding a fraction of the second component 
to the first component influence the power law approximation of the
curve. Adding 20 \% of the second component to the first component corresponds to a power law approximation
with exponent $\simeq 1.5$, while subtraction the same amount corresponds to an exponent of  $\simeq 0.5$.
As a consequence the ratio of the second component to the first component is also related to the steepness 
of the density profile, or equivalently to the concentration of mass at the center.
This analysis is confirmed by the strong correlation observed between the mass concentration (ratio of mass
to 1 scale unit to mass at 3 scale units) and the ratio between the components. As a consequence the variations
in shape of the rotation in the dBB sample are related to a variation in the concentration of mass at the center.
\section{Relation between rotation curve and and galaxy physics.}
 The rotation curves of the dBB sample are properly approximated with two PCA components, and the ratio
of these components $\eta$ is associated with the concentration of mass at the center. Let's now
investigate the relation between $\eta$ and the other galaxies parameters, magnitude $M_R$, surface brightness $\mu_R$,
scale $h_R$, mass of gas $M_G$, and total mass $M$. A significant correlation is found between the gas fraction $\log(G/M)$ and
$\eta$. No significant correlation is found between $\eta$ and other variables (see table ~\ref{table_res}). The statistical
significance of the correlation between the variables are analyzed using 2 different estimates, the Spearman correlation 
coefficient, and the a linear least-square fit (with errors in both directions). To evaluate the significance of the correlation,
 the probability that the Spearman correlation coefficient is different from zero is computed. However the problem is complicated
by the fact that each  point of a given curve has a different variance. Thus
the Spearman correlation analysis should be considered as an evaluation of the real correlation between the variable. Another
type of analysis that would take into account the different variance of the data points is to fit a straight line, and estimate
the statistical significance that the slope is different from zero. The statistical errors on the different
variable and $\eta$ are computed using the initial errors on the light curves data points from dBB. A Least square minimization
with errors in both variable is conducted (Press 2007). The chi-square of the fit is about $\simeq 3$, which is large and indicates
that other errors, probably due to to an intrinsic noise in the relation between the variables are present.
 It is assumed that this intrinsic
noise amplitude $\sigma$ is constant whatever the galaxy. The value of $\sigma$ is unknown and must be estimated from the data
. A rough estimate of $\sigma$ can be obtained by requiring that the chi-square is reduced to unity for the relevant value
 of $\sigma=\sigma_0$. The real value of $\sigma$ must be close to $\sigma_0$ but in practice $\sigma$ is unknown 
and the full space of possible realizations has to be explored, by running a large number
of numerical simulations for all values of $\sigma$. 
Among the full space of possible realizations, one has to select the sub-set of simulations
 which is consistent with the data. A simple measure of the
 consistency is that the reduced chi-square, $\chi^2$ is equal to unity when 
fitting a straight line and considering an intrinsic noise with
 amplitude $\sigma_0$. In practice the sub-set of simulations consistent with the data
 is selected using the criteria, $ \left | \chi^2-1 \right |  < 0.01$.
The statistical distribution of the slope for the sub-set is well approximated by a gaussian, with
a standard deviation $\alpha$. The statistical significance of the slope $b$ is evaluated by computing
the ratio, $b/\alpha$ (see Table ~\ref{table_res}). The association between $\eta$ and the gas fraction is very
significant, both in the Spearman correlation coefficient ($1.6$ \% chance of no association between $\eta$
 and the gas fraction), and the straight line fit (3.5 $\sigma$). There is no significant association of other 
variables with $\eta$. Is the association between $\eta$ and the gas fraction real or is it due to any systematic
effect ? One possibility would be that the association is indirect and reflects an association of each variable
with another variable. However since there is no correlation between the second variable and any other variables except $\eta$ 
 this hypothesis can be rejected. Another point is that the mass budget considered in this analysis include
also the stellar disk. Does the contribution from the stellar disk introduces a 
correlation in the variable that would be mistakenly attributed to the dark halo ? The correlation
that we observe is that galaxies with larger gas fractions are more centrally concentrated. On the
other hand the gas is converted into stars, thus a higher gas fraction is expected to corresponds
 to lower stellar content. As a consequence in galaxies with a higher gas fraction the stellar disk 
is expected to contribute less to the mass budget. Thus the effect of the stellar disk is in an
 opposite direction with respect to the observed correlation between the gas fraction and $\eta$. It is interesting to
point that the assumed correlation between higher gas fraction and lower stellar fraction ($S/M$) is confirmed
 by analyzing the dBB sample. The Pearson correlation coefficient between the stellar fraction and the gas fraction is 
is -0.53, which corresponds to a chance of only 4\% that the variable are un-correlated. Note that to derive the stellar fraction,
 the stellar mass was 
normalized by a mean statistical mass $M_S$ for the galaxy which was derived by fitting the data: $M_S \propto h_R^{1.6}$.
 This normalization was preferred to a normalization by the total mass extrapolated from the rotation curve because it
 introduces less noise. Fitting a straight line using the procedure already described for the gas fraction and $\eta$
 demonstrates that the association is significant to more than 3 $\sigma$.
 Note also that by taking into account the relation between the gas fraction and the stellar fraction and the relation
 between the gas fraction and $\eta$ one can infer the relation between the stellar fraction and $\eta$. The inferred
 slope is consistent with the data, no correlation is found between the data, but this is due to the fact that inferred
 slope is small with respect to the noise.
\begin{table}[ht]
\caption{Correlation and significance of the correlation between different
 variables and $\eta$. The lower line presents the results obtained by fitting
 a straight line to the same set of variables.}
\centering
\begin{tabular}{c c c c c c}
   & $\log(M_G/M)$ & $M_R$ & $\mu_R$ & $\log(h_R)$ & $\log(S/M)$ \\
\hline
Correlation & -0.61 & 0.03 & -0.28 & -0.18 & 0.057\\
Significance & 0.016 & 0.91 & 0.32 & 0.51 &  0.84\\
\hline
Fit Slope & -0.31 & 0.03 & -0.41 & -0.07 & -0.017 \\
$|$Slope$|$/$\sigma$ & 3.5 & 0.64 & 0.74 & 0.94 & 0.15\\
\hline
\end{tabular}
\label{table_res}
\end{table}
\begin{figure}
\centering{\epsfig{figure=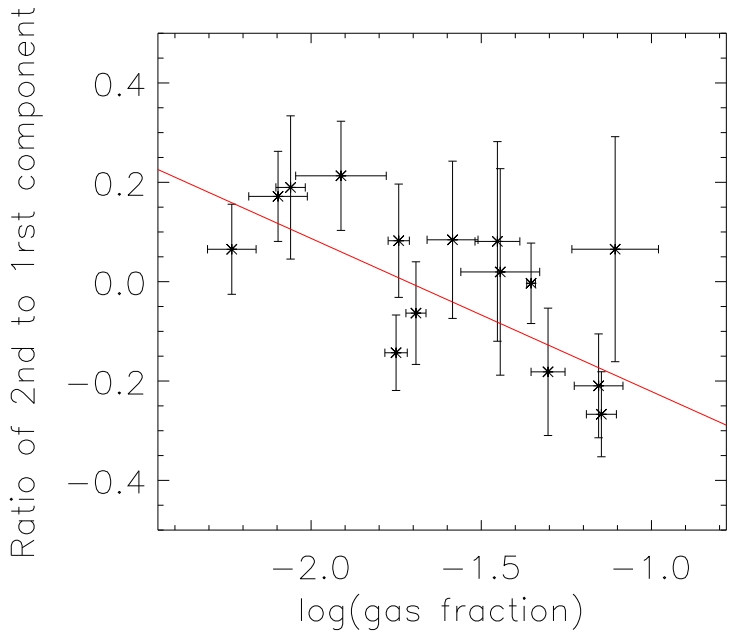,width=11cm}}
\caption{The ratio of the second to the first principal component as a function of the gas fraction.
 Note that negative values of $\eta$ corresponds to more
centrally concentrated mass profiles.}
\label{gas}
\end{figure}
\section{Discussion}
\begin{figure}
\centering{\epsfig{figure=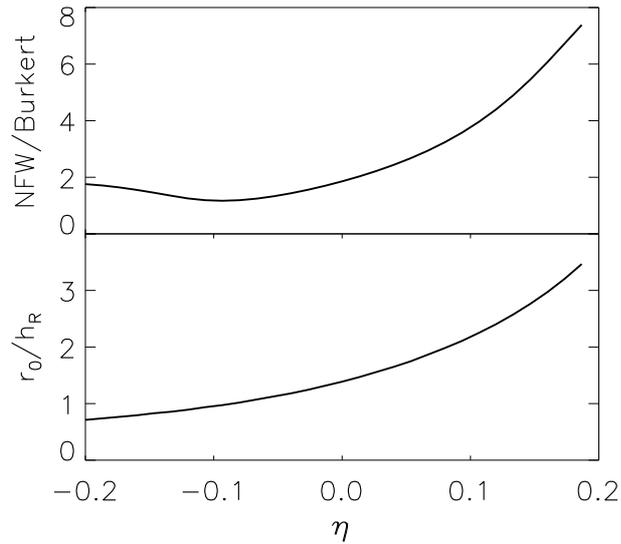,width=9cm}}
\caption{Fit of Burkert and NFW profiles to a linear combination of the first
and second components. Negative values of $\eta$ corresponds
to more centrally concentrated profiles. The lower plot presents the scale
length of the fitted Burkert profile $r_0$ as a function of $\eta$. The upper plot
presents the ratio of the residual obtained by fitting a NFW profile
to the residual obtained by fitting a Burkert profile.}
\label{burk_nfw}
\end{figure}

 The shape of galaxy rotation curve and the mass concentration at the center are related to $\eta$. Furthermore, $\eta$ 
 is related to the the gas fraction in the dBB sample. The parameter $\eta$ is also closely related
 to mass model fitting of the rotation curves (see Fig. ~\ref{burk_nfw}). Negative values of $\eta$
are associated with more concentrated mass profile, shorter Burkert scale length and better
consistency with a NFW profile when compared to positive values of $\eta$. As a consequence gas rich
galaxies will be also associated with shorter Burkert scale and will be closer to a NFW profile.
Note that this association cannot be due to a systematic bias related to the mass of the galaxies. 
One could think that more massive galaxies have steeper profile and more gas because the self
gravity is stronger and more easily overcome the supernovae winds. However since no correlation
between mass and $\eta$ is observed this systematic bias is unlikely. Actually the effect of
stronger gravity in more massive system could be compensated by more efficient star formation.   
Kim (2007) found that the gas mass fraction is correlated to the ratio
 of current to past average star formation.  The results of Kim (2007) shows
that the gas fraction is related to the evolution of galaxies, gas poor galaxies had more stellar
formation in the past than gas rich galaxies with respect to current star formation.
Statistically the gas rich galaxies are less evolved, have lower stellar fractions, and have
a better consistency with NFW profiles. 
Provided that the initial dark matter profile is close to a NFW
profile the effect of galaxy evolution via supernovae explosions is to form shallower 
profiles with nearly constant density cores (Larson 1974, Read \& Gilmore 2005, 
Governato {\it et al.} 2010). The effect of tidal encounter may boost galaxy evolution 
and influence the mass profile (Mayer {\it et al} 2001, Hayashi {\it et al} 2003). Galaxies interaction
 may also increase the star formation rate and influence
 the stellar mass function (Habergham {\it et al} 2010). Star formation, Galaxy interactions and other effects
related to galaxy evolution flattens the inner mass profile, and in the same time reduce the gas fraction
fraction in the galaxy. 
The relation between mass profile and gas fraction is thus most likely a
correlation between galaxy evolution and mass profile. 
We observe an evolutionary sequence where
galaxies with a larger gas fraction are less evolved than gas poor galaxies.
It is likely that gas poor galaxies were subject to more interactions with other galaxies,
boosting stellar formation and supernovae, while at the same time the tidal interaction has also
some influence on the mass profile. An analysis of SDSS data by Rosenbaum \& Bomans (2004) shows that gas
rich LSB forms in the voids and that some migrated to the edge of filaments while others remain
in the voids. This scenario implies that LSB galaxies had less tidal interaction than other
brighter galaxies, but the level of tidal interaction encountered by a given LSB galaxy is as
variable as the neighborhood of the galaxy is, even if they
form in relatively void regions, some encounters are still likely, and those galaxies who later fall
on more densely populated areas receive more tidal interaction than others. 
This would explain the differences of evolution in this sample, and the corresponding differences
in the rotation curves and mass profiles.
+

\begin{thebibliography}{}
%
\bibitem[]{} de Blok, W., Bosma, A., 2002, A\&A, 385, 816
\bibitem[]{} Burkert, A., 1995, ApJ, 447, L25
\bibitem[]{} Carlson, E., Machacek, M., Hall, L., 1992, ApJ, 398, 43
\bibitem[]{} Governato, F., Brook, C., Mayer, L., Brooks, A., Rhee, G., Wadsley, J., Jonsson, P., Willman, B., Stinson, G., Quinn, T., Madau, P.,  2010, Nature, 463, 203
\bibitem[]{} Habergham, S., Anderson, J., James, P., 2010, ApJ, 717, 342
\bibitem[]{} Hayashi E., Navarro J. F., Taylor J. E., Stadel J., Quinn T., 2003, ApJ, 584, 541
\bibitem[]{} Hu, W., Barkana, R., Gruzinov, A., 2000, PhRvL, 85, 1158
\bibitem[]{} Kim, Ji Hoon, Ph.D. dissertation, The star formation history of low surface brightness galaxies, 2007
\bibitem[]{} Larson, R., 1974, MNRAS, 169, 229
\bibitem[]{} Mayer L., Governato F., Colpi M., Moore B., Quinn T., Wadsley J., StadelJ., Lake G., 2001, ApJ, 547, L123
\bibitem[]{} McGaugh, S., de Blok, W., 1998, ApJ, 499, 41
\bibitem[]{} Moore, B., Gelato, S., Jenkins, A., Pearce, F., Quilis, V.,  
2000, ApJ, 535, L21
\bibitem[]{} Moore, B., Governato, F., Quinn, T., Stadel, J., Lake, G., 1998,,ApJ, 499, L5
\bibitem[]{} Moore, B., 1994, Nature, 370, 629
\bibitem[]{} Navarro, J. Frenk, C., White, S., 1997, ApJ, 490, 493
\bibitem[]{} Navarro, J. Frenk, C., White, S., 1996, ApJ, 462, 563
\bibitem[]{} Peebles, J., 2000, ApJ, 534, L127
\bibitem[]{} Press, W., Teukolsky, S., Vetterling, W., Flannery, B.,
  Numerical Recipes, Cambridge University Press, 2007
\bibitem[]{} Spergel, D., Steinhardt, P., 2000, PhRvL, 84, 3760
\bibitem[]{} Read, J., Gilmore, G., 2005, MNRAS, 356, 107
\bibitem[]{} Rosenbaum, S., Bomans, D., 2004, A\&A, 422L, 5
\bibitem[]{} Wechsler, R., Bullock, J., Primack, J., Kravtsov, A. Dekel, A.,2002, ApJ, 568, 52
%
\end{thebibliography}
\end{document}